\documentstyle[prl,twocolumn,aps,psfig]{revtex}

\def\bea{\begin{eqnarray}}
\def\eea{\end{eqnarray}}
\def\be{\begin{equation}}
\def\ee{\end{equation}}

\begin{document}
\title{On freeze-out problem in hydro-kinetic approach to A+A
collisions}
\author{Yu.M. Sinyukov$^{1,2}$, S.V. Akkelin$^{1}$ and Y. Hama$^{2}$}
\address{$^1$ Bogolyubov Inst Theor Phys, Kiev 03143, Metrologichna
14b,Ukraine,\\
$^2$ IF -- University of S\~ao Paulo, C.P. 66318, 05389-970 S\~ao
Paulo, SP, Brazil} \maketitle

\begin{abstract}
A new method for evaluating spectra and correlations in the
hydrodynamic approach is proposed. It is based on an analysis of
Boltzmann equations (BE) in terms of probabilities for constituent
particles to escape from the interacting system. The conditions of
applicability of Cooper-Frye freeze-out prescription are considered
within the method. The results are illustrated with a
non-relativistic exact solution of BE for expanding spherical
fireball as well as with approximate solutions for ellipsoidally
expanding ones.
\end{abstract}

\medskip

{\it Introduction} --- The hydrodynamic approach to multiparticle
production in hadron collisions was proposed in 1953 by Landau
\cite{landau}. It considers as the initial state a very hot and
dense thermal gas of particles soon after the collision, which
then expands hydrodynamically  until some finite time when the
picture of continuous medium is destroyed. The latter stage, so
called freeze-out, happens when the mean free path of particles
becomes comparable with the smallest of system's characteristic
dimensions: its geometrical size or hydrodynamic length.
Hydrodynamic models find serious utilization in description of
high-energy nuclear collisions, especially at CERN SPS and
Brookhaven RHIC. Studying predicted spectra of different particle
species versus the initial conditions (IC) and equation of state
(EoS), one could get information on early (partonic) stage of the
``Little Bang'', such as a possible formation of quark-gluon
plasma (QGP) or even type of the phase transition between QGP and
the hadron matter. The problem is, however, whether the predicted
spectra with given IC and EoS in hydrodynamic models are
unambiguous.

The standard and widely used method to get the spectra is the
so-called Cooper-Frye prescription (CFp) \cite{cooper}, that treat
the system at the decay stage of evolution as a locally
equilibrated ideal gas at some hypersurface. However, this
prescription presents some serious problems because,  usually, the
freeze-out hypersurface contains non-space-like sectors and, as a
result, artificial discontinuities (shock waves) across those
hypersurfaces appear to adjust CFp to energy and momentum
conservation laws \cite{Si89}. Moreover, the results of many
studies based on cascade models contradict the idea of sudden
''freeze-out'': particles escape from the system during about the
whole time of its evolution and do not demonstrate the local
equilibration at late stages. The method of continuous emission
\cite{grassi} gave an important step in the description of
particle freeze-out from 4-volume within hydrodynamic approach.
However, it is not fully satisfactory because it is not based on
Boltzmann equations (BE) and, as consequence,  when applied to
realistic systems, fails  as the particle escape probability
becomes large. In the present paper we propose an approximate
method for describing spectra within hydro-kinetic approach, which
is simpler for dense systems than pure microscopic one, and apply
it to a description of particle production from an expanding and
interacting Boltzmann gas that is initially in local equilibrium.
  
{\it Boltzmann equation, probability to escape and escaping
function} --- The BE for distribution function $f(x,p)$ in the case
of no external forces has the form:
\begin{equation}
\frac{p^{\mu}}{p^0}\frac{\partial f(x,p)}{\partial x^{\mu}}%
=F^{gain}(x,p)-F^{loss}(x,p)\,. \label{Boltzmann}
\end{equation}
The term $F^{gain}$ and $F^{loss}$ are associated with the number of
particles which respectively came to the point $(x,p)$ and leave
this point because of collisions. The term
$F^{loss}(x,p)=R(x,p)f(x,p)$ can easily be expressed in terms of the
rate of collisions of the particle with momentum $p$,
$R(x,p)=\,<\!\sigma v_{rel}\!>n(x)$. The term $F^{gain}$ has more
complicated integral structure and depends on the differential
cross-section. To develop a corresponding method, let us split the
distribution function at each space-time point into two parts:
$f(x,p)=f_{int}(x,p)+f_{esc}(x,p)$, $x=(t,{\bf x})$. The first one,
$f_{int}(x,p)$, describes the fraction of the system which will
continue to interact after the time $t$. The second one,
$f_{esc}(x,p)$, describes the particles that will never interact
after the time $t$. Of course, the latter expresses only
probabilities and cannot be ignored when one is solving BE:
possible interactions with the escaping part have also to be taken
into account, in order to find the escape probabilities.

The function $f_{esc}(x,p)$ is built up as follows. Let us denote by
$x^{\prime}\equiv(t^{\prime},{\bf x}+({\bf p}/p_0)(t^{\prime}-t))$
the space-time point where a particle in $x$ with momentum $p$ would
be, if it moved freely. Consider, at each vicinity of the
phase-space point $(x,p)$, the number of particles that have escaped
from the interacting system during the time interval
$(t^{\prime },t^{\prime }+dt^{\prime })$. This {\it additional} 
portion of escaped particles can be produced only from the 
interacting part of the system. Also, these particles are only among
those that came to the phase-space vicinity of the point 
($x^{\prime },p$) {\it just after} an interaction during the time
$dt^{\prime }\,,$ suffering the last collision there. Therefore, the
additional contribution to $f_{esc}(x,p)$ from the time interval
$(t^{\prime},t^{\prime}+dt^{\prime})$ is 
${\cal P}(x^{\prime},p)F^{gain}(x^{\prime},p)dt^{\prime}$ for
$t^{\prime}<t$ and is zero for $t^{\prime}>t$. Here ${\cal P}(x,p)$
is the probability for any {\it given} particle at $x$ with momentum
$p$ not to interact any more, propagating freely. So collecting all
the contributions starting from some initial time $t_{0}\,$, we have

\begin{equation}
f_{esc}(x,p)=f_{esc}(x_0,p)+\int\limits_{t_0}^tdt^{\prime}{\cal P}
(x^{\prime},p)F^{gain}(x^{\prime},p), \label{f-esc2}
\end{equation} 
where $x_0\equiv(t_0,{\bf x}+({\bf p}/p_0)(t_0-t))$ and
$f_{esc}(x_{0},p)$ corresponds to the portion of the system, which
is already free at $t_{0}\,$. It follows from (\ref{f-esc2}) that
\begin{equation}
\frac{1}{{\cal P}(x,p)}\frac{p^\mu}{p^0}\frac{\partial}
{\partial x^\mu}f_{esc}(x,p)=F^{gain}(x,p)\,. \label{eq-f+}
\end{equation}

The escape probability ${\cal P}(x,p)$ can be expressed explicitly
in terms of the rate of collisions along the world line of the free
particle with momentum $p$ as
\begin{equation}
{\cal P}(x,p)=\exp\left(-\int\limits_t^{\infty}dt^{\prime}
R(x^{\prime},p)\right) \label{prob-calc}
\end{equation}
or it satisfies the differential equation
\begin{equation}
\frac{1}{{\cal P}(x,p)}\frac{p^{\mu}}{p^0}\frac{\partial}
{\partial x^{\mu}}{{\cal P}(x,p)}=R(x,p)=
\frac{F^{loss}(x,p)}{f(x,p)}\,. \label{dif-esc}
\end{equation}
On the other hand, according to the probability definition,
\begin{equation}
f_{esc}(x,p)={\cal P}(x,p)f(x,p)\,. \label{prob-def}
\end{equation}
This equation, where $f_{esc}$ is given by eq. (\ref{f-esc2}) and
${\cal P}$ by eq. (\ref{prob-calc}), is one of the integral forms of
BE (\ref{Boltzmann}). One can check directly that
$f=f_{esc}/{\cal P}$ is governed by BE and, thus, our definitions of
$f_{esc}(x,p)$ and ${\cal P}(x,p)$ are consistent with BE.

For initially finite system with a short-range interaction among
particles, actually the system becomes free at large enough times
$t_{out}\,$, so ${\cal P}(x,p)\rightarrow 1$ and
$f_{esc}(x,p)\rightarrow f(x,p)$ in this limit. Our proposal is to
utilize, for the description of particle spectra in A+A
collisions, the escaping function (\ref{f-esc2}) with ${\cal
P}(x,p)$ (and thus $R(x,p)$, see eq. (\ref{prob-calc})) and
$F^{gain}$ evaluated just in the local equilibrium (l.eq.)
approximation for $f(x,p)$. In this case the  function
$f=f_{esc}/{\cal P}$ corresponds to a solution of kinetic equation
in relaxation time approximation,
\begin{equation}
\frac{p^{\mu}}{p^0}\frac{\partial f(x,p)}{\partial x^{\mu}}
=-\frac{f(x,p)-f_{l.eq.}(x,p)}{\tau (x,p)}\,, \label{relax}
\end{equation}
with the relaxation time $\tau(x,p)$ being expressed through the
rate of collisions $R_{l.eq.}(x,p)$ of a particle with momentum
$p,\,\tau=1/R_{l.eq.}\,$. When $t \rightarrow \infty$, the 
relaxation time $\tau \rightarrow \infty$ also, indicating a 
transition to the free streaming regime. The eq. (\ref{relax}) can 
be derived also from BE with approximation of r.h.s. by 
$F^{gain}=R_{l.eq.}\,f_{l.eq.}\,$ and $F^{loss}=R_{l.eq.}\,f$ 
\cite{Liboff}. Note that the macroscopic parameters of l.eq. 
distribution function are governed, in general, by non-ideal 
hydrodynamic equations that are derived from eq. (\ref{relax}) in 
the standard way (see, e.g., \cite{Liboff}). 

The approximation proposed is based on collecting all the
{\it liberated} particles during the system evolution and,
therefore, takes into account rather non-equilibrium processes,
similar to particle emission from the periphery of a system.
Obviously, it should describe well the spectra in a fast transition
of the system from an initial equilibrium state to free streaming
(e.g., explosion), as well as in a fairly smooth transition when
l.eq. is nearly maintained till rather low densities. Formally,
comparing with the exact solution of BE, we neglect in eq.
(\ref{f-esc2}) the term
\begin{equation}
{\cal P}F^{gain}-{\cal P}_{l.eq.}R_{l.eq.}\,f_{l.eq.}\,,
\label{deviat}
\end{equation}
integrated along the world line of a particle from $t_{0}$ to
$t_{out}\,$. At part of a trajectory crossing sufficiently dense
space-time region, neither $F^{gain}$ nor ${\cal P}$ differ much
from the corresponding hydro terms in eq. (\ref{deviat}). At the
later stage of expansion, when the densities are small (normally 
$n(x)\sim t^{-3}$), the values of $F^{gain}$ terms, both for hydro 
and exact solution, are rather small. So the integrated contribution
of the neglected term, eq. (\ref{deviat}), coming from that part of
a trajectory, is also small no matter how much the l.eq.
distribution function $f_{l.eq.}$ deviates there from the exact
solution $f$. As for the most delicate transition region, one can
expect that the neglected term (\ref{deviat}) in this region could
not be important, too. In details, the transition from initial l.eq.
state to free streaming is rather fast at a periphery of the system,
so the correspondent integrated contribution of (\ref{deviat}) is
small. The contribution of the neglected term from trajectories
crossing a central part of the system could be, for l.eq. initial
distributions, also small if the deviation from l.eq. is slight
until rather small densities. We shall discuss below some examples
of such a behavior.

{\it Particle spectra and correlations} --- To describe the
inclusive spectra of particles
\begin{equation}
p^0\frac{dN}{d{\bf p}}=\langle a_p^+a_p\rangle \,,\
p_1^0p_2^0\frac{dN}{d{\bf p{\tt _{1}}}d{\bf p}{\tt _{2}}}=
\langle a_{p_1}^+a_{p_2}^+a_{p_1}a_{p_2}\rangle \,,
\label{spectra-def}
\end{equation}
the asymptotic equality $f_{esc}(x,p)=f(x,p)$ can be used, replacing
the total distribution function $f$ in all irreducible averages in
(\ref {spectra-def}),
\begin{equation}
\langle a_{p_1}^+a_{p_2}\rangle =\int_{\sigma_{out}}d\sigma_{\mu}
p^{\mu}\exp{(iqx)}f(x,p)\,, \label{average-wigner}
\end{equation}
by $f_{esc}\,.$ Here, $p=(p_{1}+p_{2})/2\,$, $q=p_{1}-p_2$ and the
hypersurface $\sigma _{out}$ just generalizes $t_{out}\,$. Applying
the Gauss theorem and recalling that
$\partial_{\mu}[p^{\mu} exp{(iqx)}]=0$ for particles on mass shell,
one obtains, using respectively general equations (\ref{eq-f+}) and
(\ref{Boltzmann}) and supposing their analytical continuation off
mass shell,
\begin{eqnarray}
\langle a_{p_1}^+a_{p_2}\rangle&=&p^{\mu}\int_{\sigma_0}
d\sigma_{\mu}f_{esc}(x_0,p)e^{{iqx}} \nonumber \\
&+&p^0\int_{\sigma_0}^{\sigma_{out}}\!d^4x\,{\cal P}(x,p)
F^{gain}(x,p)e^{{iqx}}\,,\hspace*{-0.3cm} \label{sp-e} \\
\langle a_{p_1}^+a_{p_2}\rangle&=&p^{\mu}\int_{\sigma_0}d\sigma_\mu
f(x_{0},p)e^{{iqx}}  \nonumber \\
&+&p^{0}\!\!\int_{\sigma_{0}}^{\sigma _{out}}\hspace*{-0.3cm}
d^{4}x(F^{gain}(x,p)\!-\!F^{loss}(x,p))e^{{iqx}}\,. \label{sp-f}
\end{eqnarray}

Thus, the use of escaping function as the asymptotic interpolation
to the solution of BE is equivalent to taking,
 as the source function for the spectra and correlations, the
4-volume emission function $S={\cal P}F^{gain}$  together with
direct emission $f_{esc}(x_{0},p)$  from an initial 3D
hypersurface $\sigma_0$. The CFp, defined by eq.
(\ref{average-wigner}) with substitution
$\sigma_{out}\!\rightarrow\sigma_{f.o.}$ and $f\!\rightarrow
f_{l.eq.}$, treats particle spectra as results of rapid {\it
conversion} of a l.eq. hadron system into a gas of free particles
at some hypersurface $\sigma_{f.o.}$. Formally, it corresponds to
 taking the cross-section tending to infinity at
$t<t_{\sigma_{f.o.}}$ (to keep system in l.eq. state)  and zero
beyond $t_{\sigma _{f.o.}}$. Then ${\cal P}(t,{\bf
x},p)=\theta(t-t_{\sigma_{f.o.}}\!({\bf x}))$ (and so $f_{esc}=0$
at $t<t_{\sigma _{f.o.}}$), and $S={\cal
P}F^{gain}=\delta(t-t_{\sigma_{f.o.}}\!({\bf x}))f_{l.eq.}$ in eq.
(\ref{sp-e}).

Let us now analyze the particle-spectrum formation process based
on simple analytical models. It was recently shown \cite{Akkelin}
that, if at an initial time $t=t_{0}=0\,$ a non-relativistic ideal
gas has an ellipsoidally symmetric Gaussian density distribution
and a self-similar velocity profile ${\bf u}(x)$, then the
solution for ideal hydrodynamics has the form (${\bf C}= {\bf
v}-{\bf u}(x),{\bf v}={\bf p}/m,V=\prod_{i=1}^{3}X_{i}$)
\begin{equation}
f(t,{\bf x},
{\bf v})=\frac{N}{V}(\frac{m}{(2\pi)^{2}T})^{\frac{3}{2}}\!\exp
(-\frac{m{\bf C}^{2}}{2T}-\!\sum_{i=1}^{3} \frac{x_{i}^{2}}{2X_{i}^{2}})
\label{ex-s}
\end{equation}
where
\begin{equation}
X_{i}\stackrel{..}{X}_{i}=\frac{T}{m}\,,\ \ T=T_{0}\left(
\frac{V_{0}}{V}%
\right) ^{2/3}\!,\ \ u_{i}=\frac{\stackrel{.}{X}_{i}}{X_{i}}x_{i}\ .
\label{id-sol}
\end{equation}

In the spherically symmetric case, when all $X_{i}$ are equal, the
l.eq. distribution function (\ref{ex-s}) with (\ref{id-sol}) is an
exact solution of BE \cite{Csorgo}. One can easily check that the
momentum spectrum as well as the interferometry radius, computed at
any time $t$ (at any isotherm) or at any other hypersurface
enclosing the system, are identical to those calculated at the
initial time $t=0$. The reason is that, in the spherically
symmetric case, the l.eq. function (\ref{ex-s}) makes the l.h.s. of
BE (\ref {Boltzmann}) zero, as well as the r.h.s., rendering the
evolution of the {\it interacting} gas similar to a free streaming.
Then the volume integral in eq. (\ref{sp-f}) vanishes and in this
case CFp formally gives the correct spectrum but its physical
meaning is completely different from the naively expected one. There
is no clearly defined unique freeze-out hypersurface: neither the
escape probability ${\cal P}$ nor the emission function $S$ reveal a
sharp behavior in ($t,{\bf x}$) plane, but are rather smooth
functions. This is demonstrated in Fig. 1 with the average (in
${\bf v}$)\hfilneg\  

\begin{figure}[tbp]
\vspace*{8.8cm}
\begin{center}
\includegraphics{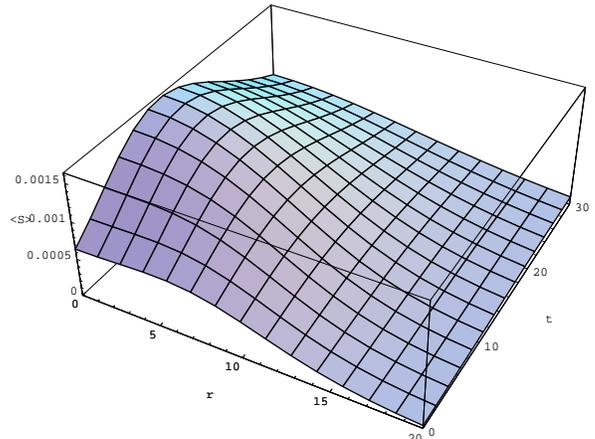}
\end{center}
\vspace{-4.3cm} \caption{The space-time $(t,r)$ dependence of the
emission function averaged over momenta for an expanding
spherically symmetric fireball containing 400 particles with mass
1 GeV and with cross section $\sigma=40$ mb, initially at rest and
localized  with Gaussian radius parameter $R=7$ fm and temperature
$T_{0}=0.130$ GeV.}
\vspace*{-.cm}
\end{figure}

\noindent emission function
$<\!S\!>=<\!{\cal P}F^{gain}\!>=<\!{\cal P}Rf\!>$, where ${\cal P}$
is calculated according to eq. (\ref{prob-calc}).

Therefore, we can conclude that the difference between final
spectra (and interferometry radii) at $\sigma_{out}$ and what
could be found at $\sigma_0$ is due to dissipative effects
(deviations from l.eq.), which make the integral over 4-volume in
eq. (\ref{sp-f}) non vanishing. The contribution of dissipative
effects can be essential even if the evolution of the system is
governed with good accuracy by ideal fluid hydrodynamics, that
take place for fairly high densities or/and cross-sections, and so
$f(x,p)\approx f_{l.eq.}(x,p)$.  In this case, $p^\mu\partial_\mu
f(x,p)\approx p^\mu\partial_\mu f_{l.eq.}(x,p) \sim\kappa
f_{l.eq.}(x,p)$,  where $\kappa$ does not depend on the density
and the cross section  but tied, in particular, with symmetry
properties of the hydrodynamic expansion. For the above discussed
exact spherically symmetric solution of BE, $\kappa =0$, but in
the general case $\kappa\neq0$ and, as a result, for high
densities particle rescattering can lead to a serious changing of
momentum spectra as compared to the initial ones, depending on the
initial conditions.

Typically, however, hadron system is not at local equilibrium
during fairly long later stage of evolution and $f$ could largely
deviate from $f_{l.eq.}\,$. The proposed method of escape
probabilities with l.eq. approximation for ${\cal P}(x,p)$ and
$F^{gain}$ allows to use hydrodynamic approach for calculations of
spectra even in the situation when finite inhomogeneous systems go
through all stages: from local equilibrium to free streaming. Let
us illustrate this, based on the exact solution (\ref {id-sol}) of
ideal hydrodynamics for an initially compressed ellipsoid in the
longitudinal direction, containing the same gas as in the previous
example. The results are presented in Fig. 2, where
single-particle spectra at $t=8$ fm for particles that ``escaped''
(became free according to eq.(\ref{f-esc2})) up to this moment are
shown versus those corresponding to frozen-out ones according to
CFp, applied to l.eq. distribution function\hfilneg\

\begin{figure}[tbp]
\vspace*{6.7cm}
\begin{center}
\includegraphics{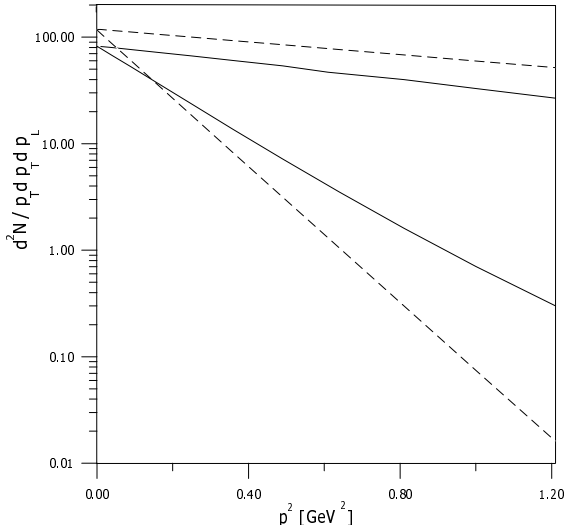}
\end{center}
\vspace{-0.4cm} \caption{The transverse (bottom, at $p_L=0$) and
longitudinal (top, at $p_T=0$) spectra of particles, escaped until 
$t=8$ fm/c from an expanding ellipsoidally symmetric fireball of the 
same particles as in Fig.1, initially at rest and localized  with 
Gaussian radius parameters $X_1=X_2=7\,$fm, $X_3=0.7\,$fm and 
temperature $T_0=0.300\,$GeV. Dashed lines correspond to spectra 
calculated according to CFp, applied to l.eq. distribution function 
at $T(t)=0.063$ GeV.} 
\end{figure} 

\noindent
(\ref{ex-s}) on the isotherm $T(t)$. One can see that the
effective temperature of the transverse spectrum calculated by using
$f_{esc}(x,p)$ is higher than the one given by the l.eq.
distribution function. This is in agreement with the results of
partonic cascade algorithm of ref. \cite{Gyulassy} and also with
those of continuous emission \cite{grassi}. The longitudinal
spectrum shows an opposite tendency.

The solution of the ideal hydrodynamics (\ref{ex-s}), (\ref{id-sol})
also shows that
$(\stackrel{.}{X}_i/X_i)/(\stackrel{.}{X}_j/X_j)\rightarrow 1$
with time increas\-ing, so the velocity field of the expanding
system tends to a spherically symmetric one. Such a tendency of the
velocity field is preserved, in the central region, also by the
solutions of Navier-Stokes equation with the same initial
conditions, if transport coefficients are calculated according to
Chapman-Enskog (ChE) method (we used a hard-sphere model of
interaction). Because of this, the deviation from l.eq. in the
central part of the system, calculated within ChE method, is rather
negligible until the density becomes quite small. Then, there is a
hope to apply CFp for the soft-momentum spectra, since these
particles are mainly emitted from there. As for the hard-momentum
spectra, one expects that such particles are mostly radiated from
the periphery of the system at the early times, because of large
hydrodynamic velocities and fast transition to free streaming there.
So, in a rough approximation, one can apply the generalized CFp
taking into account first, $p$-dependent hypersurface
$\sigma_{f.o.}$ and, second, the deviations on $\sigma_{f.o.}$ of
the distribution function from l.eq. due to dissipative effects.

{\it Conclusion} --- The proposed method allows to describe, in
hydrodynamic approach for $A+A$ collisions, the evolution of the
matter from l.eq. till free streaming. The method differs from the
continuous emission developed in refs. \cite{grassi}, where the
central object is the interacting component $f_{int}$ which is
approximated by the l.eq. distribution function. However it is
possible to show, by using an exact solution of BE, that $f_{int}$
is neither isotropic nor thermal distribution in its local rest
frame, even at a moderate value of $\langle{\cal P}(x,p)\rangle$.

It is worth to note that our method, applied to the evolution of
rarefied hadron gas, overlaps with transport models like RQMD. In
this aspect, the exact solution of BE discussed above is useful as
a test of numerical cascade algorithms. The advantage of the method
of escaped particles reveals when the system is not dilute, has more
complicated interactions and displays a collective behaviour. The
hydrodynamics is still (or even more) suitable for the description
of such a system, and so our method based on calculations of
probabilities for constituent particles to escape can be used,
considering a concrete model for the particle interaction with
surrounding medium.

The analysis of particle-liberation process demonstrates the crucial 
role of dissipative effects in formation of one- and two-particle 
spectra. These effects should be taken into account when experimental 
data from SPS and RHIC are treated in hydrodynamic approach. 

{\it Acknowledgments }--- We are grateful to T. Osada for helping in
making the graphs, and P. Braun-Munzinger for fruitful discussions.
This work has been supported by the FAPESP grants 00/02244-7 and
01/05122-0, S\~ao Paulo, Brazil, by German-Ukrainian Grant No.
2M/141-2000, French-Ukrainian CNRS Grant No. Project 8917.

\end{document}